\documentclass[aps,prl,twocolumn, superscriptaddress, amsmath]{revtex4-1}

\usepackage[usenames,dvipsnames]{color}
\usepackage{graphicx}
\usepackage{dcolumn}
\usepackage{simplewick}
\usepackage{bm}
\usepackage{amsmath}
\usepackage{braket}
\usepackage{float}
\usepackage{threeparttable, tablefootnote}
\usepackage{hyperref}
\usepackage{natbib}
\usepackage{multirow}
\usepackage{booktabs}
\usepackage{adjustbox}
\usepackage{animate}
\usepackage[percent]{overpic}
\usepackage{subcaption}
\usepackage{hyperref}
\hypersetup{colorlinks=true, urlcolor=blue, citecolor=blue}

\begin{document}
\title{Negative and Zero Linear Compressibility in MCN (M = Ag, Au, Cu): A First-Principles Study}
\author{Arlies Valdespino}
\email{arliesv@usf.edu}
\affiliation{Department of Physics, University of South Florida, Tampa, Florida 33620, USA}
\author{Abduljelili Popoola}
\affiliation{Department of Physics, University of South Florida, Tampa, Florida 33620, USA. Currently at the School of Mechanical and Materials Engineering, Washington State University, Pullman, Washington, 99164, USA.}
\author{Sergey Lisenkov}
\affiliation{Department of Physics, University of South Florida, Tampa, Florida 33620, USA}
\author{Inna Ponomareva}
\affiliation{Department of Physics, University of South Florida, Tampa, Florida 33620, USA}

\begin{abstract}
Negative linear compressibility (NLC) is the counterintuitive phenomenon in which a crystallographic axis expands under hydrostatic pressure. A related phenomenon, zero linear compressibility (ZLC), occurs when an axis shows no length change under pressure. Both responses are rare and arise in materials with highly anisotropic mechanical properties. 
Motivated by recent reports of anomalous behavior in metal cyanides—including negative thermal expansion and NLC—we use first-principles calculations to investigate the potential of the MCN family (M = Ag, Au, Cu) to exhibit ZLC or NLC. Our simulations reveal three main findings. First, we predict the existence of previously unreported phases: the P6mm phase for AgCN and CuCN, and the R3m phase for AuCN. Second, all six members of the MCN family studied exhibit extreme elastic anisotropy, which indeed leads to ZLC or NLC in each case. Third, we identify a mechanism for these responses distinct from previously reported ones. The mechanism arises from the unique ``bamboo forest" geometry of the material: weakly interacting rigid rods whose sparse packing allows pressure to be accommodated through changes in packing density rather than rod compression.
 This mechanism enables the anomalous response to persist over a wide pressure range, contrasting favorably with other materials that exhibit similar behavior. Our findings expand the relatively small family of materials known to exhibit ZLC or NLC and provide deeper insight into the microscopic origin of these unusual mechanical responses.
\end{abstract}

\maketitle
\section*{Introduction}
Under hydrostatic pressure, most materials contract uniformly in all spatial directions. However, a counterintuitive mechanical behavior—NLC—has been observed in a small but growing number of solids, wherein one or more crystallographic axes expand under isotropic compression. These anomalous responses have attracted attention not only for their fundamental scientific interest, but also for their potential applications in pressure sensors, actuators, and mechanically tunable metamaterials \cite{Baughman1998,Goodwin2008}. 

The phenomenon of NLC is rare and often subtle in magnitude, with known examples spanning framework materials (e.g., MIL-53 and Ag$_3$[Co(CN)$_6$]), inorganic oxides (e.g., TeO$_2$ and MgF$_2$), and molecular crystals (e.g., methanol monohydrate)~\cite{Cairns2015}. One of the prevailing mechanisms to enable NLC is a hinging or ``wine-rack'' motif, in which a network of rigid units rotates or pivots under pressure, causing one dimension to elongate even as the overall volume decreases~\cite{Ogborn2012}. Additional mechanisms reported in the literature include tilting of polyhedral units, bond bending, soft interlayer interactions (e.g., layer sliding~\cite{Zeng2017}), and empty-channel structural motifs~\cite{Colmenero2021}. Despite this conceptual progress, achieving strong, stable NLC across a relatively wide pressure range remains challenging. Many candidate materials suffer from mechanical instability, phase transitions under modest pressure, or very weak NLC magnitudes \cite{Cairns2015}. 

An equally intriguing phenomenon is ZLC, in which a crystallographic direction exhibits almost no change in length under hydrostatic pressure. Such behavior is often associated with highly anisotropic bonding environments, where extremely stiff structural motifs resist compression along specific directions while softer modes accommodate the overall volume reduction~\cite{Jiang2018}. Materials exhibiting ZLC are of considerable technological interest because they provide exceptional dimensional stability under external stress, making them promising candidates for precision mechanical components, pressure-resistant optical devices, and resonators~\cite{Jiang2018, Yu2020, Zeng2022}. 
Despite recent advances, materials that exhibit robust ZLC or NLC remain scarce, and the known microscopic mechanisms that give rise to these phenomena are still few in number.

In this context, the metal cyanides MCN—where M = Ag (R3m), Au (P6mm), and Cu (R3m and C222$_1$)—have been reported to exhibit negative thermal expansion  along the $c$ direction \cite{Hibble2010, PhysRevB.93.134307} which makes them potential candidates for ZLC or NLC. Indeed, the coexistence of negative thermal expansion  and NLC observed in other cyanides suggests that the MCN family may exhibit a similar relationship \cite{Goodwin2008}. Moreover, recent high-pressure XRD and Raman spectroscopy revealed giant NLC ($-20.5$~TPa$^{-1}$) in the low-temperature orthorhombic C222$_1$ phase of CuCN, with a hinging mechanism identified as the driver \cite{PhysRevLett.134.126102}. However, to date there are limited reports of NLC/ZLC in the R3m and/or P6mm phases of these materials. A previous computational study has reported the existence of NLC in AgCN (R3m) and nearly ZLC in AuCN (P6mm) \cite{Korabelnikov2021}. 

In this paper, we use first-principles simulations to explore MCN family in order to address the aforementioned challenges in the area of anomalous mechanical responses.  In particular, the aims of our study are  (i) to predict three new phases in the MCN family: P6mm phase for AgCN and CuCN, and the R3m phase for AuCN; (ii) to predict that these phases along with the experimentally reported phases  exhibit either ZLC or NLC, thereby expanding the known family of such materials; (iii) to  trace the atomistic origin of these rare properties to extreme elastic anisotropy; and (iv) to identify a mechanism for ZLC/NLC in these materials that differs from the wine-rack, tilting, and sliding mechanisms previously reported. Our findings advance fundamental understanding of ZLC and NLC phenomena while expanding the family of candidate materials for future fundamental  and applied research.

\section{Computational Methodology}

The crystal structures of AgCN in the R3m phase and AuCN in the P6mm phase were obtained from Bowmaker et al.~\cite{Bowmaker1998}, who determined them using powder neutron diffraction. The R3m structure for CuCN was obtained from the Materials Project database~\cite{Horton2025, Jain2013}. In addition, we considered P6mm phases of AgCN and CuCN and the R3m phase for AuCN, which have not been reported experimentally so far. These phases were obtained by replacing the metal site in the experimentally reported structures. Thus, our set of structures consists of three cyanides (AgCN, AuCN, and CuCN) in two phases (R3m and P6mm), that is six structures altogether.

The structures were computationally studied using density functional theory (DFT) as implemented in the Vienna Ab-initio Simulation Package (VASP)~\cite{Kresse1993, Kresse1996_CMS}. To identify the most reliable functional for our study, we tested six exchange-correlation functionals: the Perdew–Burke–Ernzerhof (PBE) parametrization~\cite{Perdew1996, Perdew1997} with and without the D3 dispersion correction of Grimme et al.~\cite{Grimme2010}, the modified PBE parametrization for solids (PBEsol)~\cite{PhysRevLett.100.136406} also with and without the D3 correction, and the meta-GGA functional r$^2$SCAN~\cite{Furness2020} with and without D3 corrections.
Ion–electron interactions were modeled using the projected augmented wave (PAW) potentials from the VASP pseudopotential library. Technical details on the energy cutoff and k-point mesh are provided in Supplemental Material Table ~\ref{tab:computational_parameters}. 

Full structural relaxations were carried out until forces on the ions were less than 0.01 eV/\AA.  For self-consistent calculations, the convergence criterion was set to 10$^{-7}$~eV. Linear compressibilities, Young's moduli, and bulk moduli were computed from single-crystal elastic constants using the finite difference method implemented in VASP and analyzed using the ELATE software~\cite{Gaillac_2016}.

\section{Structural and elastic  properties of MCN }

\begin{figure*}
    \centering

    \begin{subfigure}[b]{0.42\textwidth}
        \centering
        \caption{R3m}
        \label{fig:AgCN_r3m}
        \includegraphics[width=\textwidth, trim=70 40 40 10]{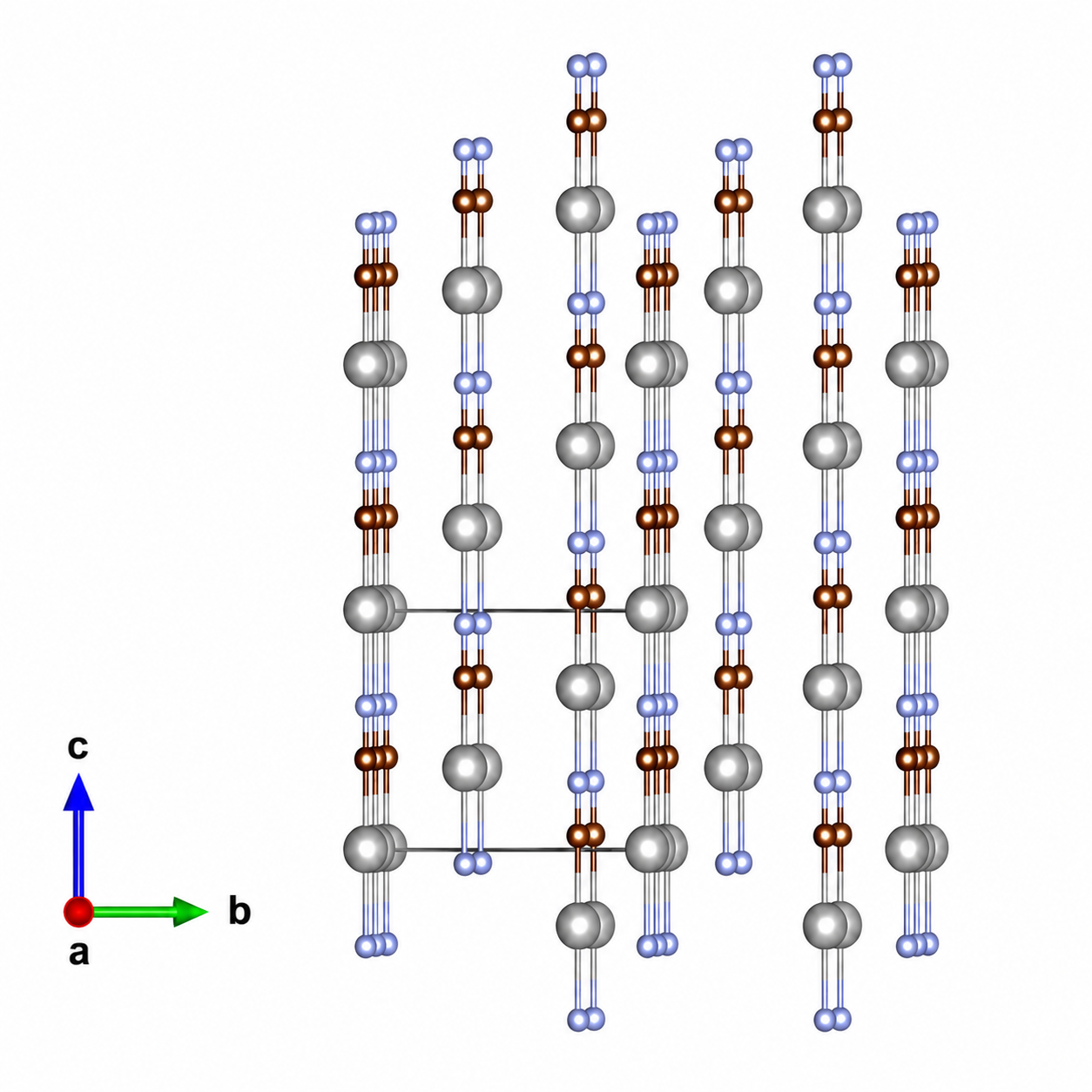}
    \end{subfigure}
    \hspace{0.06\textwidth}
    \begin{subfigure}[b]{0.29\textwidth}
        \centering
        \caption{P6mm}
        \label{fig:AgCN_p6mm}
        \includegraphics[width=\textwidth, trim=70 40 40 10]{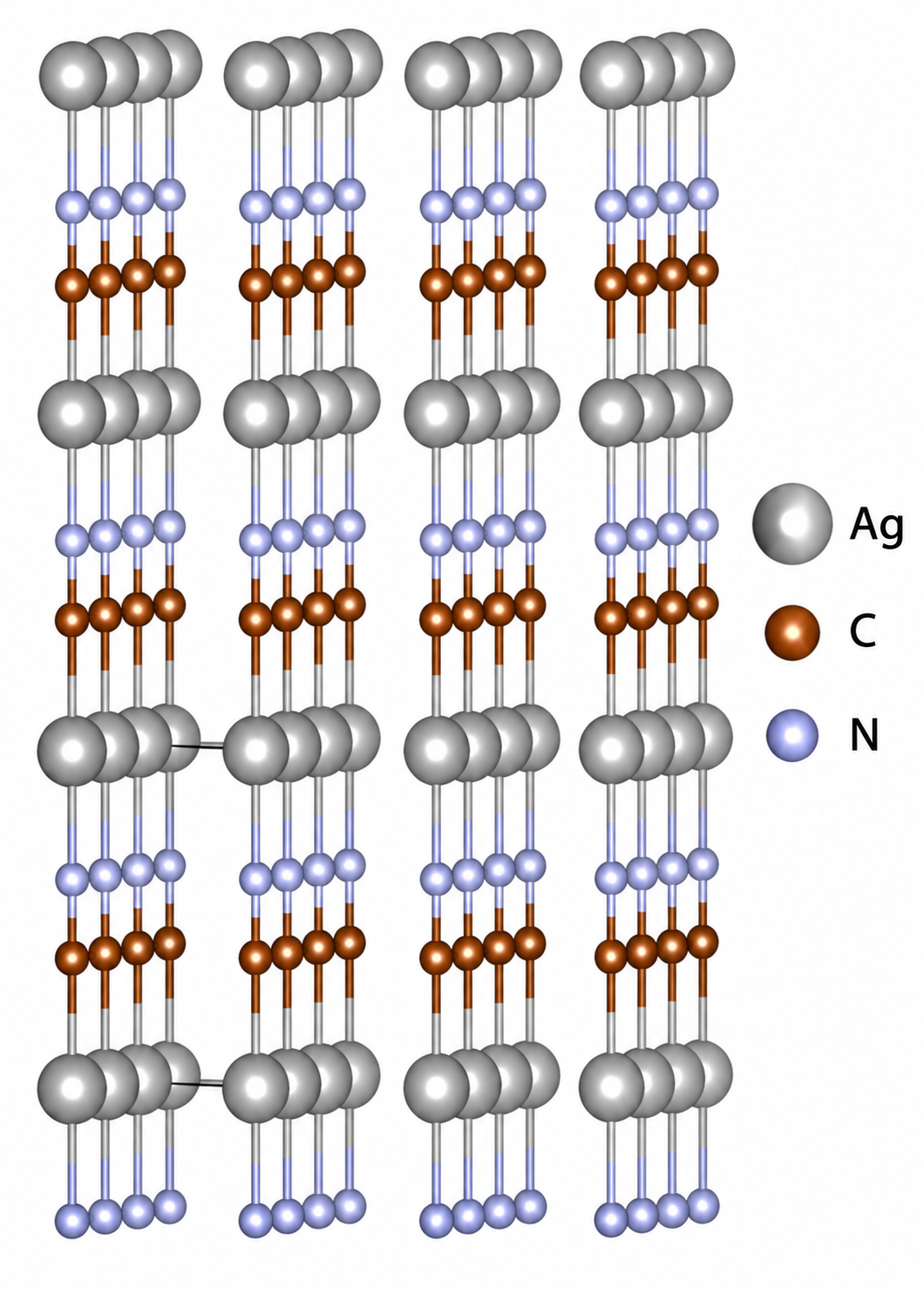}
    \end{subfigure}

    \caption{Structures of AgCN in R3m and P6mm phases. }
    \label{fig:AgCN_phases}
\end{figure*}

We performed full structural relaxation for all six structures using the six exchange-correlation functionals indicated in the previous section. In many instances, the R3m and P6mm structures relaxed into the Cm and Cmm2 phases, respectively. However, when the tolerance for symmetry determination was increased from $10^{-5}$ \AA\ to $10^{-3}$ \AA, the original R3m and P6mm symmetries were restored. Given that the latter tolerance is more relevant to experimental resolution, we chose to continue with the R3m and P6mm phases. These phases are visualized in Fig.~\ref{fig:AgCN_phases} and can be viewed as consisting of MCN strings assembled in two different ways. The resulting structure is reminiscent of a ``bamboo forest" with relatively sparse packing in the $ab$ plane. Note that the energy difference between the lower- and higher-symmetry phases is less than $10^{-5}$ eV, which is small and justifies our approximation. Table~\ref{tab:lattice_constants_compact} gives lattice parameters computed with different exchange-correlation functionals for the phases that have experimental data available, providing a comparison between computational and experimental results. We find the best agreement for the r$^2$SCAN and PBE-D3 functional, followed closely by PBEsol (less than 1.8\% error). We note that dispersion corrections improve predictions for PBE but not for the other functionals. 

\begin{table*}
  \centering
  \caption{Lattice parameters (\AA) for experimentally available phases of MCN computed from DFT, along with experimental values from Refs.~\cite{Bowmaker1998,Reckeweg2003}. Percent errors are given in parentheses.}
  \label{tab:lattice_constants_compact}
  \begin{ruledtabular}
  \begin{tabular}{llcccccccc}
    Compound & Axis & PBE & PBE-D3 & PBEsol & PBEsol-D3 & r$^2$SCAN & r$^2$SCAN-D3 & Exp. \\
    \hline
    \multirow{2}{*}{AgCN (R3m)}
      & $a$ & 6.99 (17\%) & 6.03 (0.5\%) & 6.11 (1.8\%) & 5.75 (-4.2\%) & 5.98 (-0.3\%) & 5.79 (-3.5\%) & 6.00 \\
      & $c$ & 5.22 (-0.8\%) & 5.24 (-0.4\%) & 5.18 (-1.5\%) & 5.18 (-1.5\%) & 5.25 (-0.2\%) & 5.25 (-0.2\%) & 5.26 \\
      \hline
    \multirow{2}{*}{AuCN (P6mm)}
      & $a$ & 3.78 (11\%) & 3.50 (2.9\%) & 3.36 (-1.2\%) & 3.27 (-3.8\%) & 3.45 (1.5\%) & 3.35 (-1.5\%) & 3.40 \\
      & $c$ & 5.08 (-0.2\%) & 5.08 (-0.2\%) & 5.06 (-0.6\%) & 5.05 (-0.8\%) & 5.08 (-0.2\%) & 5.07 (-0.4\%) & 5.09 \\
      \hline
    \multirow{2}{*}{CuCN (R3m)}
      & $a$ & 6.89 (17\%) & 5.91 (0.0\%) & 5.97 (1.0\%) & 5.58 (-5.6\%) & 5.94 (0.5\%) & 5.69 (-3.7\%) & 5.91 \\
      & $c$ & 4.81 (-1.0\%) & 4.81 (-1.0\%) & 4.78 (-1.6\%) & 4.78 (-1.6\%) & 4.80 (-1.2\%) & 4.79 (-1.4\%) & 4.86 \\
  \end{tabular}
  \end{ruledtabular}
\end{table*}

The relative energies for all phases considered here are reported in Table~\ref{tab:stable_phase}. In each case, the experimentally relevant phase is the reference. For AuCN, all functionals predict that the experimentally observed P6mm phase is lower in energy than our proposed R3m phase. For AgCN, the predictions disagree on the relative energetic ordering of the two phases. Since the energy difference between the two phases is relatively small and its sign is functional-dependent, we conclude that the phases are energetically competitive. For CuCN, all functionals except PBE-D3 predict that our proposed P6mm phase is lower in energy than the experimentally reported R3m phase. Again, this fact, in conjunction with the relatively small energy differences, leads us to conclude that the two phases are energetically competitive.

Based on this analysis, we will proceed with investigations of all six phases, but we will limit ourselves to the three exchange-correlation functionals that perform best in predicting lattice constants: PBE-D3, PBEsol, and r$^2$SCAN.

\begin{table*}
\centering
\caption{Energy difference (meV per formula unit) relative to the experimentally stable phase indicated in bold.}
\label{tab:stable_phase}
\begin{ruledtabular}
\begin{tabular}{lcccccc}
Structure & PBE & PBE-D3 & PBEsol & PBEsol-D3 & r$^2$SCAN & r$^2$SCAN-D3 \\
\hline

AgCN (P6mm) & 11.1 & 22.00 & -48.8 & -9.62 & 8.13 & -3.01 \\
\textbf{AgCN (R3m)} & 0.00 & 0.00 & 0.00 & 0.00 & 0.00 & 0.00 \\

\hline

\textbf{AuCN (P6mm)} & 0.00 & 0.00 & 0.00 & 0.00 & 0.00 & 0.00 \\
AuCN (R3m) & 44.9 & 114.65 & 167.0 & 172.12 & 123.0 & 166.30 \\

\hline

CuCN (P6mm) & -3.01 & 1.89 & -38.3 & -14.56 & -17.37 & -21.40 \\
\textbf{CuCN (R3m)} & 0.00 & 0.00 & 0.00 & 0.00 & 0.00 & 0.00 \\

\end{tabular}
\end{ruledtabular}
\end{table*}

We next assess the elastic properties of these materials (Table \ref{tab:elastic_combined}) using these three exchange-correlation functionals. All experimentally relevant structures were found to be mechanically stable. We note that, in some instances, calculations predicted negative shear elastic constants, which led to  mechanical instability. These values were revised by carrying out direct simulations in which small shear strains (ranging from -0.6\% to 0.6\%) were applied to the structure and the associated stresses were computed. The linear slope of the stress-strain relationship was used to obtain more accurate elastic constants, resulting in mechanically stable structures. The  exceptions are  AuCN in R3m phase with PBEsol and r$^2$SCAN functionals. 
However, PBE-D3 and PBEsol-D3 predict the AuCN R3m phase to be mechanically stable, suggesting that dispersion corrections stabilize the structure. We therefore retain this phase in our investigation.

In all cases, the $C_{33}$ elastic constant exceeds $C_{11}$ by an order of magnitude, revealing strong elastic anisotropy — a key requirement for zero or negative compressibility~\cite{Cairns2015}. Furthermore, the maximum Young's modulus is very close to $C_{33}$, indicating both that the $c$-direction is the stiffest and that the perpendicular plane experiences little relaxation. The bulk modulus is significantly closer to $C_{11}$ than to $C_{33}$, suggesting that it is dominated by the soft response in the $ab$ plane.

The minimum linear compressibility, also reported in Table~\ref{tab:elastic_combined}, occurs along the $c$-direction for most materials, with the exception of the R3m phase of CuCN computed with the PBEsol functional, where unusually large negative compressibility (-49.10~TPa$^{-1}$) is predicted. However, given that PBEsol is found to be less reliable for this study, we do not explore this finding further. 
Most values are very small in magnitude, indicating near-zero compressibility along the $c$-direction, consistent with the exceptionally large elastic constant in that direction. Negative compressibility is predicted for AgCN in the R3m phase with all functionals but PBE-D3. For other materials, the sign of compressibility in R3m phase depends on the choice of functional. In P6mm phase compressibility remains positive. Nevertheless, in all instances the values are very small, predicting nearly ZLC for the entire family.

\begin{table*}
\setlength{\tabcolsep}{3pt}
\caption{Elastic constants $C_{ij}$ (GPa) and derived elastic properties for CuCN, AgCN, and AuCN. $E_{\text{max}}$ (GPa) is the maximum Young's modulus, which occurs along the $c$ direction. $\beta_{\text{min}}$ (TPa$^{-1}$) is the minimum linear compressibility, also along the $c$ direction unless noted otherwise (* implying it does not). $K$ (GPa) is the bulk modulus from the Hill averaging scheme. $E_{max}$ entries marked with ``-" correspond to unphysical values and therefore not reported. Experimentally stable phases are shown in bold.}

\label{tab:elastic_combined}
\begin{ruledtabular}
\begin{tabular}{llcccccccccc}
Mat. & XC & $C_{11}$ & $C_{12}$ & $C_{13}$ & $C_{14}$ & $C_{33}$ & $C_{44}$ & $C_{66}$ & $E_{\max}$ & $\beta_{\min}$ & $K$ \\
\hline

\multirow{3}{*}{\textbf{AgCN (R3m)}}
 & PBE-D3 & 21.14 & 7.27 & 12.60 & 0.003 & 406 & 0.96 & 6.91 & 395 & 0.29 & 35.63 \\
 & PBEsol & 9.60 & 3.72 & 11.02 & 1.05 & 449 & 0.68 & 2.91 & 431 & -1.52 & 32.20 \\
 & r$^2$SCAN & 15.70 & 6.76 & 15.16 & 0.58 & 399 & 0.33 & 4.40 & 379 & -0.93 & 33.60 \\
\hline

\multirow{3}{*}{AgCN (P6mm)}
 & PBE-D3 & 17.11 & 5.12 & 1.81 & - & 456 & 1.42 & 6.00 & 456 & 1.84 & 33.67 \\
 & PBEsol & 17.54 & 5.49 & 5.53 & - & 538 & 0.47 & 6.04 & 536 & 0.97 & 39.42 \\
 & r$^2$SCAN & 18.00 & 6.30 & 4.05 & - & 480 & 1.16 & 5.85 & 479 & 1.39 & 36.30 \\
\hline

\multirow{3}{*}{AuCN (R3m)}
 & PBE-D3 & 23.01 & 7.33 & 6.75 & 1.90 & 670 & 0.82 & 7.91 & 667 & 0.82 & 49.61 \\
 & PBEsol & 5.60 & 1.77 & 4.42 & -1.83 & 653 & -1.02 & 1.89 & - & -0.37 & 39.90 \\
 & r$^2$SCAN & 12.30 & 4.55 & 6.69 & -2.01 & 638 & -0.79 & 3.71 & - & 0.33 & 43.00 \\
\hline

\multirow{3}{*}{\textbf{AuCN (P6mm)}}
 & PBE-D3 & 23.46 & 8.43 & 8.04 & - & 685 & 1.03 & 7.46 & 681 & 0.73 & 51.34 \\
 & PBEsol & 22.21 & 9.61 & 14.93 & - & 789 & 3.05 & 6.33 & 775 & 0.08 & 58.65 \\
 & r$^2$SCAN & 21.80 & 9.10 & 12.40 & - & 713 & 1.94 & 6.30 & 703 & 0.28 & 53.50 \\
\hline

\multirow{3}{*}{\textbf{CuCN (R3m)}}
 & PBE-D3 & 17.49 & 5.92 & 8.83 & -0.21 & 579 & 0.14 & 5.76 & 573 & 0.41 & 42.59 \\
 & PBEsol & 7.34 & 2.89 & 8.85 & 1.07 & 613 & 0.52 & 2.23 & 597 & -49.10* & 39.69 \\
 & r$^2$SCAN & 18.10 & 7.00 & 11.20 & -0.79 & 595 & 1.17 & 5.50 & 585 & 0.19 & 44.60 \\
\hline

\multirow{3}{*}{CuCN (P6mm)}
 & PBE-D3 & 15.12 & 3.72 & 1.93 & - & 602 & 1.22 & 5.73 & 602 & 1.32 & 40.64 \\
 & PBEsol & 18.67 & 4.84 & 5.51 & - & 685 & 0.82 & 6.93 & 682 & 0.78 & 47.72 \\
 & r$^2$SCAN & 16.70 & 5.03 & 4.16 & - & 639 & 0.66 & 5.88 & 637 & 0.97 & 44.20 \\
\end{tabular}
\end{ruledtabular}
\end{table*}

We note that the values of linear compressibility computed above correspond to the zero-pressure limit. We next investigate whether near-zero or negative compressibility persists over a wider pressure range. To this end, we applied hydrostatic pressures of up to 13~GPa and computed the structural response of the MCN family with r$^2$SCAN and PBE-D3 functionals. The corresponding evolution of the unit-cell volume is shown in Fig. \ref{fig:volume_diff} of the Supplemental Material. In all systems, the volume decreases smoothly with increasing pressure, indicating the absence of pressure-induced structural instabilities over the investigated pressure range. The data for the relative change in lattice constants computed with r$^2$SCAN functional are shown in Fig.~\ref{fig:lattice_df_all} while data computed with PBE-D3 functional are given in Supplemental Material Fig.~\ref{fig:lattice_diff_pbe}. The figure confirms a highly anisotropic response of the structures to hydrostatic pressure. In all cases, we find a significant decrease in the $a$ lattice parameter and a very minimal change in the $c$ lattice parameter. This strong anisotropy allows the necessary volume decrease to be accommodated almost entirely through the contraction of the $a$ lattice parameter, owing to the one-dimensional connectivity of the structure. This could be visualized in a ``bamboo forest'' picture: rigid rods (the MCN chains) that are loosely packed in the $ab$ plane can move closer together under pressure without compressing the rods themselves, leaving the $c$ direction largely unaffected (or even expanded, as seen for AgCN).

The insets in each panel show zoomed-in dependencies of the $c$ lattice parameter on pressure, where we observe an increase in $c$ as a function of pressure for AgCN in the R3m phase. The PBE-D3 functional does not predict NLC (see Supplemental Material Fig. \ref{fig:lattice_diff_pbe}); however, the value, while positive, is very low. Thus, our predictions are consistent in that respect.

\begin{figure*}
    \centering
    \includegraphics[width=\linewidth]{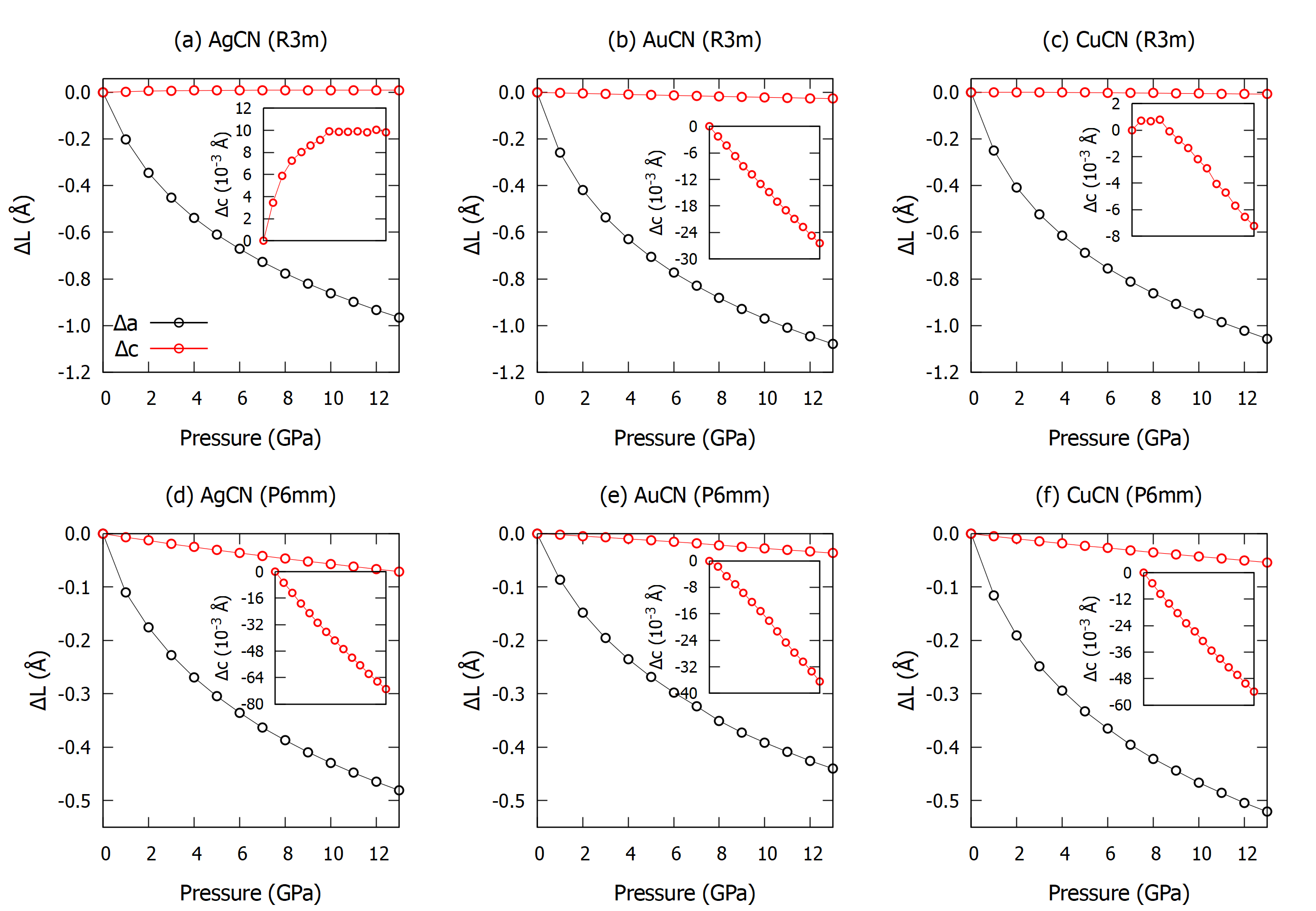}
    \caption{Change in lattice parameters as a function of pressure computed using r$^2$SCAN. Insets show a zoomed-in view of $\Delta c$.}
    \label{fig:lattice_df_all}
\end{figure*}

\begin{figure*}
    \centering
    \includegraphics[width=\linewidth]{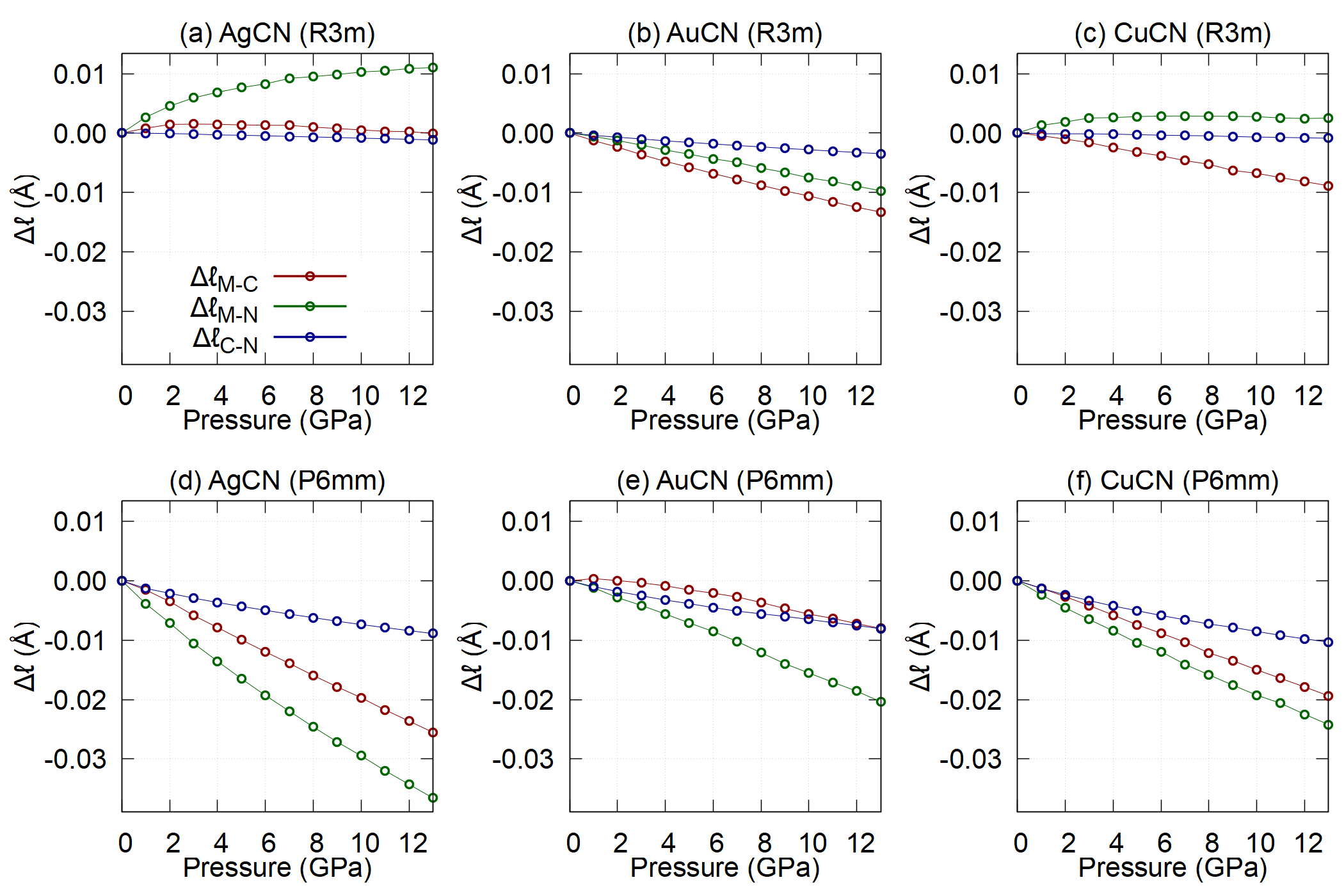}
    \caption{Difference in bond lengths with respect to  zero pressure structure along $c$ direction for M-C/N and C-N bonds.}
    \label{fig:BL_comp}
\end{figure*}

To further understand the origin of ZLC and NLC along the $c$ direction, we computed interatomic distances between neighboring atoms as a function of pressure using structures obtained with r$^2$SCAN functional; the results are presented in Fig.~\ref{fig:BL_comp}. The data reveal that for the R3m phase of AgCN, there is no change in the Ag--C and C$\equiv$N bond lengths, while the Ag--N bond length increases, giving rise to NLC in this case.  Likewise, the Cu--N bond length increases in the R3m phase of CuCN as a function of pressure, while the C$\equiv$N bond length does not change. However, the decrease in the Cu--C bond length prevents this material from exhibiting NLC, likely due to differences in metal--carbon bonding between Cu and Ag. For the remaining structures, all three bond lengths decrease with pressure, though only slightly (at most $-1.8\%$). In the P6mm phase, the M--N bond (where M denotes the metal) is the softest for all materials, yet the overall anisotropic response remains small, resulting in near-zero rather than negative linear compressibility.

For further insight, we computed phonon dispersion curves for all six structures using the finite difference method as implemented in VASP with the r$^2$SCAN and PBE-D3 functional. There was no significant difference in the phonon dispersions between different materials in the same phase. The phonon dispersions for both phases computed with r$^2$SCAN are shown in Fig.~\ref{fig:AgCN_AuCN_phonons}. Data for PBE-D3 are given in Supplemental Material Fig. \ref{fig:AgCN_AuCN_phonons}. The data reveal the existence of phonons with extremely high frequencies (up to 68~THz with r$^2$SCAN), which is rather rare and indicates extremely stiff bonds involving light atoms. Previously, phonon dispersion curves for cyanides below around 65~THz have been computationally reported in Ref.~\onlinecite{PhysRevB.93.134307}, and they agree with ours. Experimental data from Refs.\cite{Bowmaker1998} are added to Fig. ~\ref{fig:AgCN_AuCN_phonons} and reveal excellent agreement between computational and experimental data further validating our methodology. 

Let us now recall that phonon frequency is proportional to  $\sqrt{k/m}$,  where $k$ and $m$ are the spring constant and atomic mass, respectively,  so that stiff bonds contribute to a high force constant $k$, while light atoms decrease the denominator $m$. The insets in Fig.~\ref{fig:AgCN_AuCN_phonons} indicate the directions of atomic displacements in the high-frequency modes, while Fig. \ref{fig:side_by_side_animations} of the Supplemental Material provides an animation. The high-frequency modes are associated with vibrations of light atoms along the one-dimensional chains of the material. The one-dimensional nature of the structure confines electron sharing and/or redistribution to one direction, resulting in the formation of extremely strong bonds. 
Furthermore, the mode associated with ionic vibrations perpendicular to the chain is only 1.2~THz (approximately 56 times smaller than the high-frequency mode of 67~THz in AgCN), indicating extremely soft bonding between the chains, which enables easy compression.

Another interesting feature of the phonon dispersion is the presence of a very large gap (between approximately 15 and 67 THz for AgCN and between approximately 18 and 68 THz for AuCN). Materials with a large phonon gap are actively pursued due to their promise in heat flow control, phononics, and thermoelectricity. An additional phonon gap exists between approximately 8 and 15 THz for AgCN and between approximately 11 and 18 THz for AuCN. Moreover, the  branches are quite flat, which is consistent with the one-dimensional connectivity of the material. Flat bands give rise to ultra-low group velocities for phonons and are likely to lead to ultra-low and highly anisotropic lattice thermal conductivity~\cite{PICHANUSAKORN201019}.

Thus, our calculations reveal that the origin of near zero or even negative compressibility in MCN family in R3m and P6mm phases is the 1D connectivity of the structure where extremely rigid  M-C-N rods are held together by relatively weak interactions. We believe that this is an original mechanism that has not been proposed in the literature so far and therefore advances fundamental understanding of the incompressibility. 
\begin{figure*}
    \centering

    \begin{subfigure}[b]{0.49\textwidth}
        \centering
        \begin{overpic}[width=\textwidth]{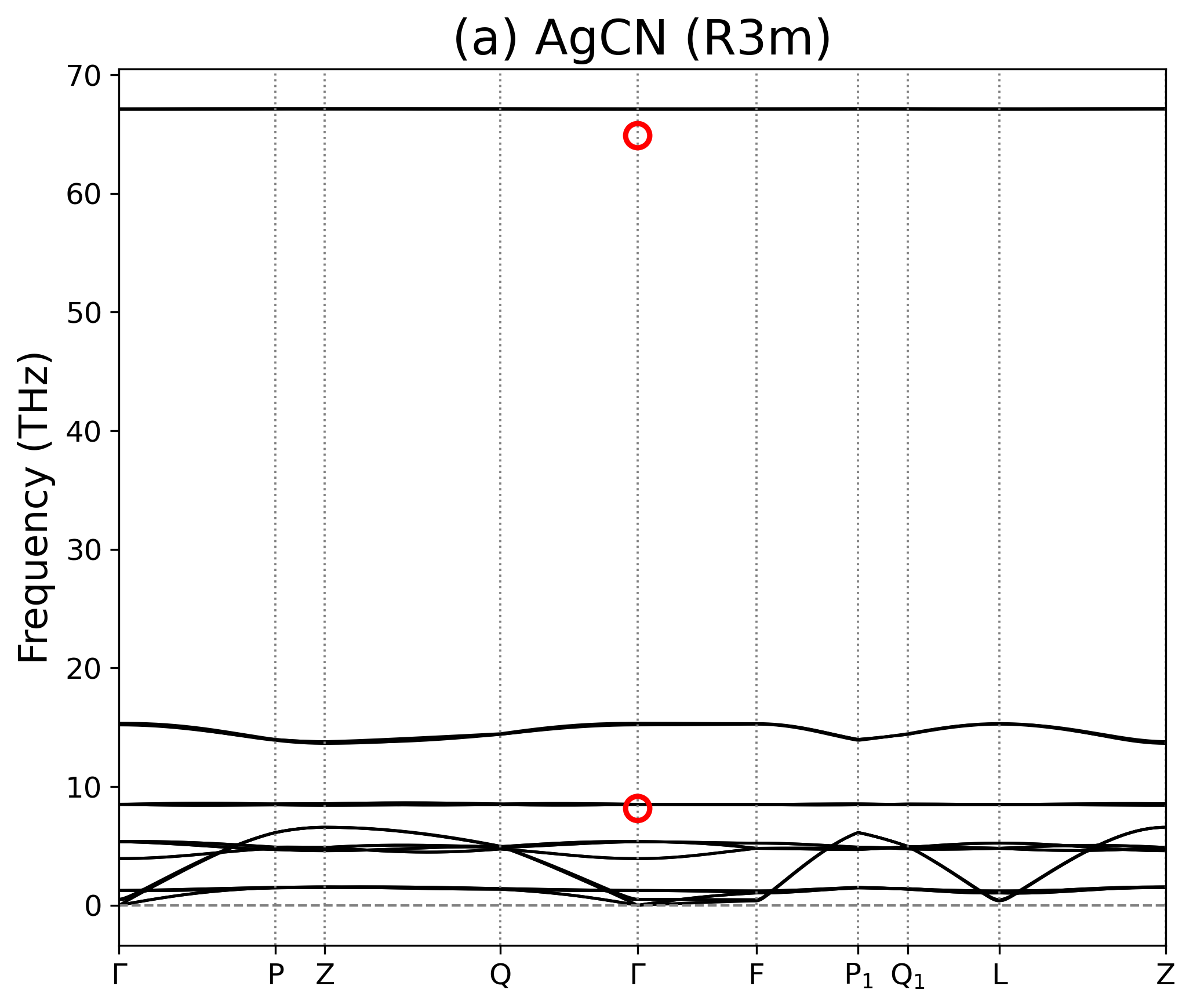}
            \put(58,36){%
                \includegraphics[width=0.3\textwidth]{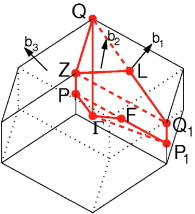}
            }
            \put(-30,20){%
                \includegraphics[width=1.3\textwidth]{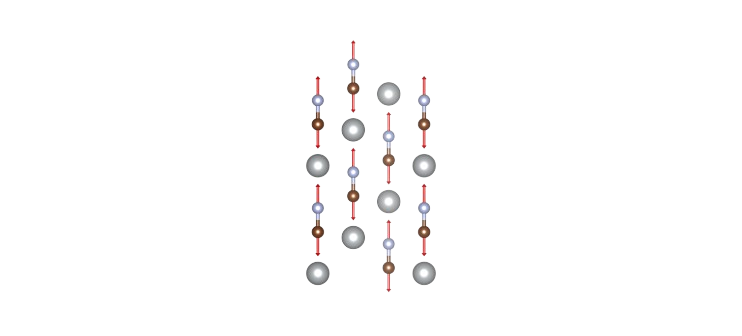}
            }
        \end{overpic}
        \label{fig:AgCN_phonons}
    \end{subfigure}
    \hfill
    \begin{subfigure}[b]{0.49\textwidth}
        \centering
        \begin{overpic}[width=\textwidth]{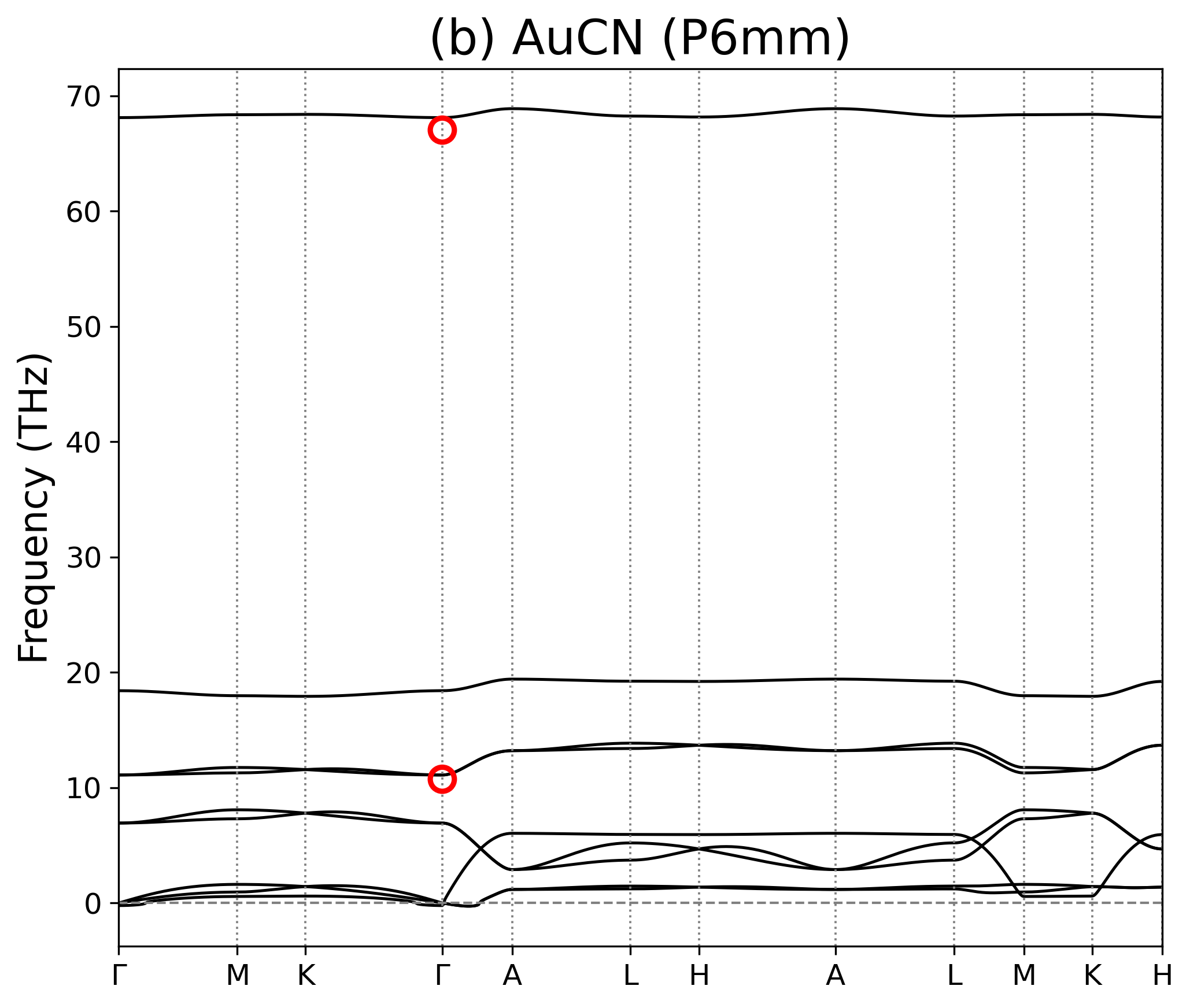}
            \put(57,45){%
                \includegraphics[width=0.4\textwidth]{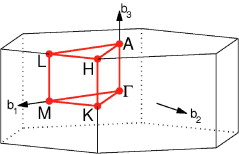}
            }
            \put(-35,26){%
                \includegraphics[width=1.3\textwidth]{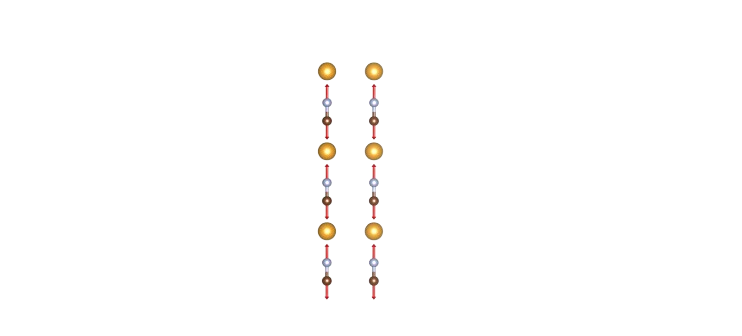}
            }
        \end{overpic}
        \label{fig:AuCN_phonons}
    \end{subfigure}

    \caption{Phonon band structures of  the R3m phase of AgCN (a) and  the P6mm phase of AuCN (b) computed using r$^2$SCAN. Insets on the right show the corresponding high-symmetry paths used in the phonon calculations (adapted from \cite{SETYAWAN2010299}). Insets on the left display the displacement vectors for the high-frequency modes. Red circles indicate the experimental high-frequency stretching and low-frequency bending modes from Ref.~\cite{Bowmaker1998}.
    }
    \label{fig:AgCN_AuCN_phonons}
\end{figure*}

\section{conclusions}

In summary, we have used first-principles simulations to investigate the R3m and P6mm phases of AgCN, AuCN, and CuCN using six exchange-correlation functionals: PBE, PBE-D3, PBEsol, PBEsol-D3, r$^2$SCAN, and r$^2$SCAN-D3. By comparison with experimental data for lattice parameters, we find that r$^2$SCAN and PBE-D3 perform best in this study followed closely by PBEsol. Our simulations predict previously unreported phases — P6mm for AgCN and CuCN, and R3m for AuCN — which are energetically competitive   with  their experimentally reported counterparts. All phases were found to exhibit highly anisotropic elastic properties, with elastic constants along the $c$-direction being an   order of magnitude larger than those in the $ab$ plane. This anisotropy results in near  ZLC or even NLC along the $c$-direction in the zero-pressure limit. We extended our predictions to pressures up to 13~GPa and found that these anomalous phenomena persist over this wide pressure range. It should be noted that we have not considered the possibility of phase transitions, as we are not aware of competing phases, except for C222$_1$ which has partial occupancies, making it challenging for computational investigation. However, from comparison of enthalpy differences at different pressures between the two phases considered here we found that for CuCN r$^2$SCAN predicts R3m phase to become more favorable than P6mm at 11~GPa pressure, while with PBE-D3 R3m starts off favorable, but P6mm becomes favorable at 1~GPa (see Fig. \ref{fig:enthalpy_diffs}(e)-(f) of Supplemental Material).   
The anomalous response to pressure originates from a highly anisotropic mechanical response, where the structure is nearly incompressible along the direction of the MCN rods and very soft in the perpendicular plane due to the sparse packing of the chains — a ``bamboo forest" geometry. At the atomistic level, the chains derive their rigidity from extremely strong bonds connecting light elements, as revealed by our phonon calculations. The rods interact with each other relatively weakly, as inferred from the low frequency of phonon modes associated with in-plane displacement of the rods. We believe that our findings both contribute to the family of materials that exhibit rare ZLC and NLC and deepen understanding of these phenomena by identifying a novel mechanism responsible for the anomalous responses.

\section{Acknowledgment}

This work was supported by the U.S.
Department of Energy, Office of Basic Energy Sciences, Division of Materials Sciences and Engineering under Grant No. DE-SC0005245.  Computational support was provided by the National Energy Research Scientific Computing Center (NERSC), a U.S. Department of Energy, Office of Science User Facility located at Lawrence Berkeley National Laboratory, operated under Contract No. DE-AC02-05CH11231 using NERSC award BES-ERCAP-0025236.

\newpage
\onecolumngrid
\begin{center}
   \textbf{\Large Supplemental Material}
\end{center}

\setcounter{table}{0}
\renewcommand{\thetable}{S\arabic{table}}

\setcounter{figure}{0}
\renewcommand{\thefigure}{S\arabic{figure}}

\begin{figure}[h!]
    \centering

    \begin{subfigure}[b]{0.48\textwidth}
        \centering
        \animategraphics[width=\textwidth,loop,autoplay]{30}{AgCN_images/AgCN-}{0}{85}
        \caption{AgCN (R3m)}
        \label{fig:AgCN_ani}
    \end{subfigure}
    \hfill
    \begin{subfigure}[b]{0.48\textwidth}
        \centering
        \animategraphics[width=\textwidth,loop,autoplay]{30}{AuCN_images/AuCN-}{0}{85}
        \caption{AuCN (P6mm)}
        \label{fig:AuCN_ani}
    \end{subfigure}

    \caption{Side-by-side animations of high frequency stretching modes for AgCN (R3m) and AuCN (P6mm).}
    \label{fig:side_by_side_animations}
\end{figure}

\begin{table}[h!]
\centering
\caption{Plane-wave energy cutoffs (ENCUT) and $\Gamma$-centered k-point meshes used for the DFT calculations of the MCN systems and their modified structures.}
\label{tab:computational_parameters}
\begin{tabular}{lccc}
\hline
\textbf{Structure} & \textbf{Functionals} & \textbf{ENCUT (eV)} & \textbf{$\Gamma$-centered k-mesh} \\
\hline

AgCN (R3m) 
& PBE/D3, PBEsol/D3, r$^2$SCAN 
& 800 
& $9 \times 9 \times 7$ \\

AgCN (P6mm) 
& PBE/D3, PBEsol/D3, r$^2$SCAN 
& 800 
& $8 \times 8 \times 4$ \\

AuCN (P6mm) 
& PBE/D3, PBEsol/D3, r$^2$SCAN 
& 900 
& $10 \times 10 \times 6$ \\

AuCN (R3m)
& PBE/D3, PBEsol/D3, r$^2$SCAN 
& 900 
& $9 \times 9 \times 6$ \\

CuCN (R3m) 
& PBE/D3, PBEsol/D3, r$^2$SCAN 
& 600 
& $9 \times 9 \times 7$ \\

CuCN (P6mm) 
& PBE/D3, PBEsol/D3, r$^2$SCAN 
& 800 
& $10 \times 10 \times 5$ \\

All R3m
& r$^2$SCAN-D3
& 800
& $7 \times 7 \times 7$ \\

All P6mm
& r$^2$SCAN-D3
& 800
& $13 \times 13 \times 7$ \\

\hline
\end{tabular}
\end{table}

\begin{figure}[htbp]
    \centering
    \includegraphics[width=\linewidth]{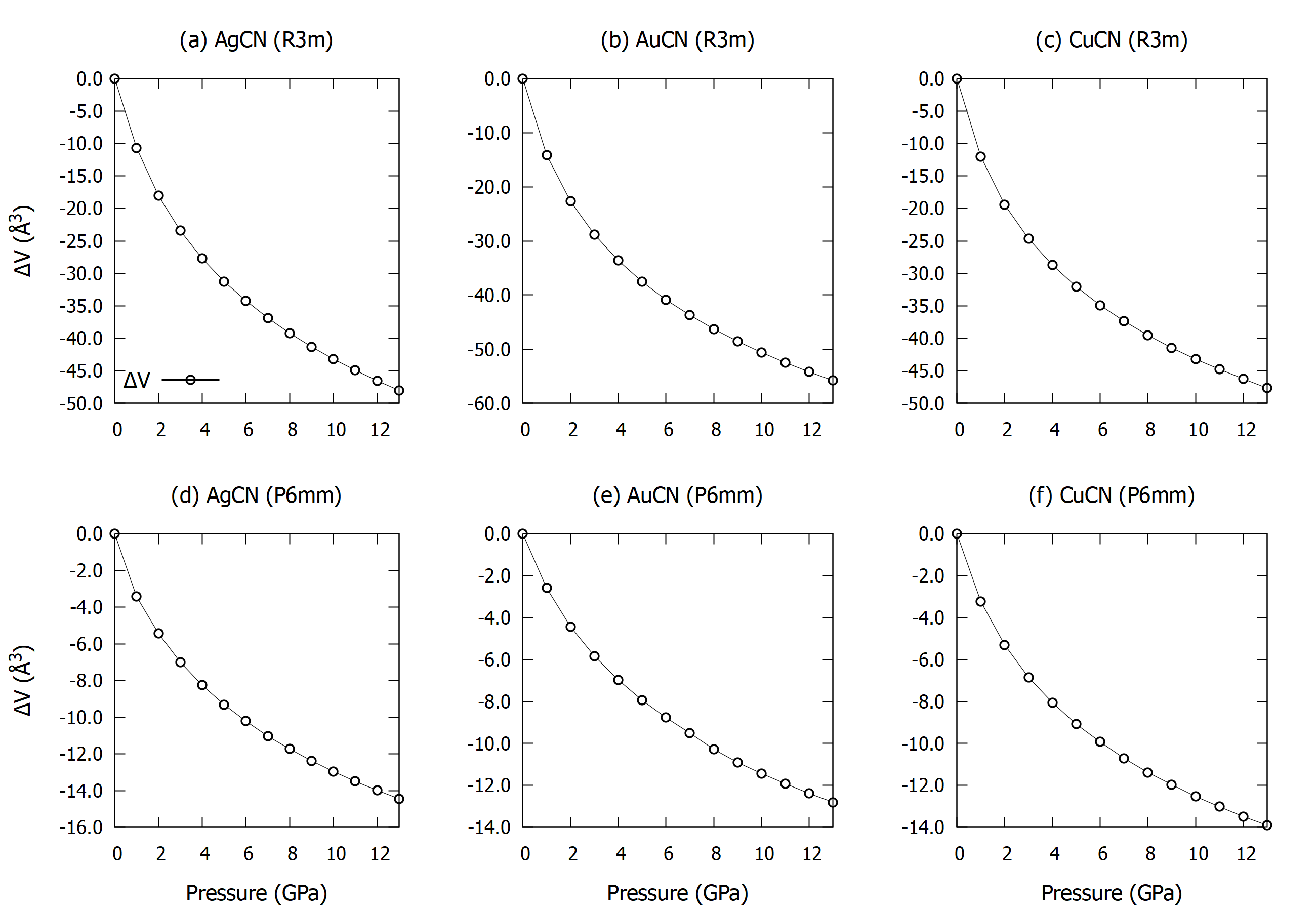}
    \caption{Change in unit-cell volume as a function of hydrostatic pressure for the studied MCN structures using the r$^2$SCAN functional. The volume change is computed relative to the zero-pressure structure, $\Delta V = V(P)-V(0)$.}
    \label{fig:volume_diff}
\end{figure}

\begin{figure}[htbp]
    \centering
    \includegraphics[width=\linewidth]{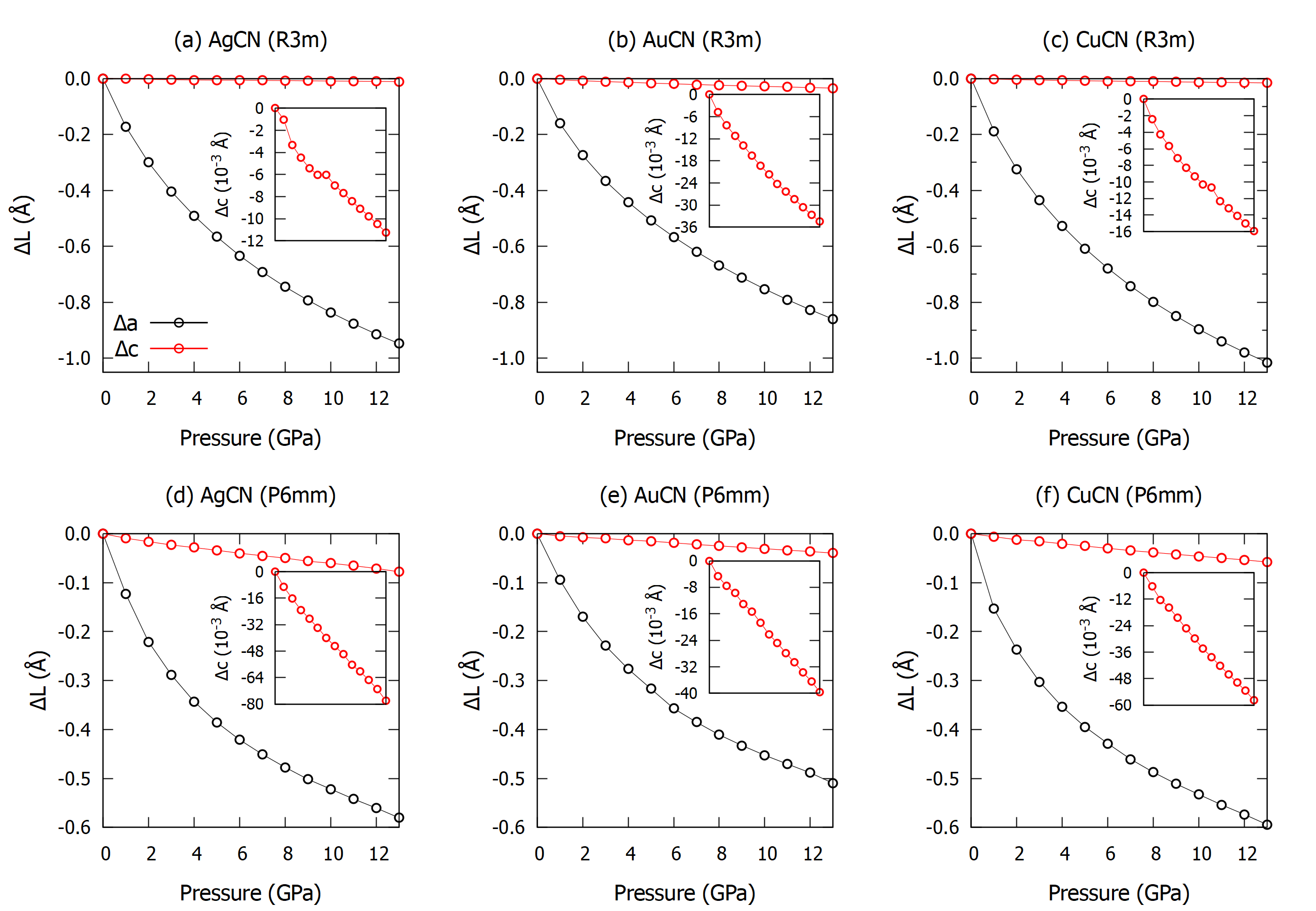}
    \caption{Change in lattice parameters as a function of pressure using PBE-D3. Insets show a zoomed-in view of $\Delta c$.}
    \label{fig:lattice_diff_pbe}
\end{figure}

\begin{figure}[h]
    \centering

    \begin{subfigure}[b]{0.49\textwidth}
        \centering
        \begin{overpic}[width=\textwidth]{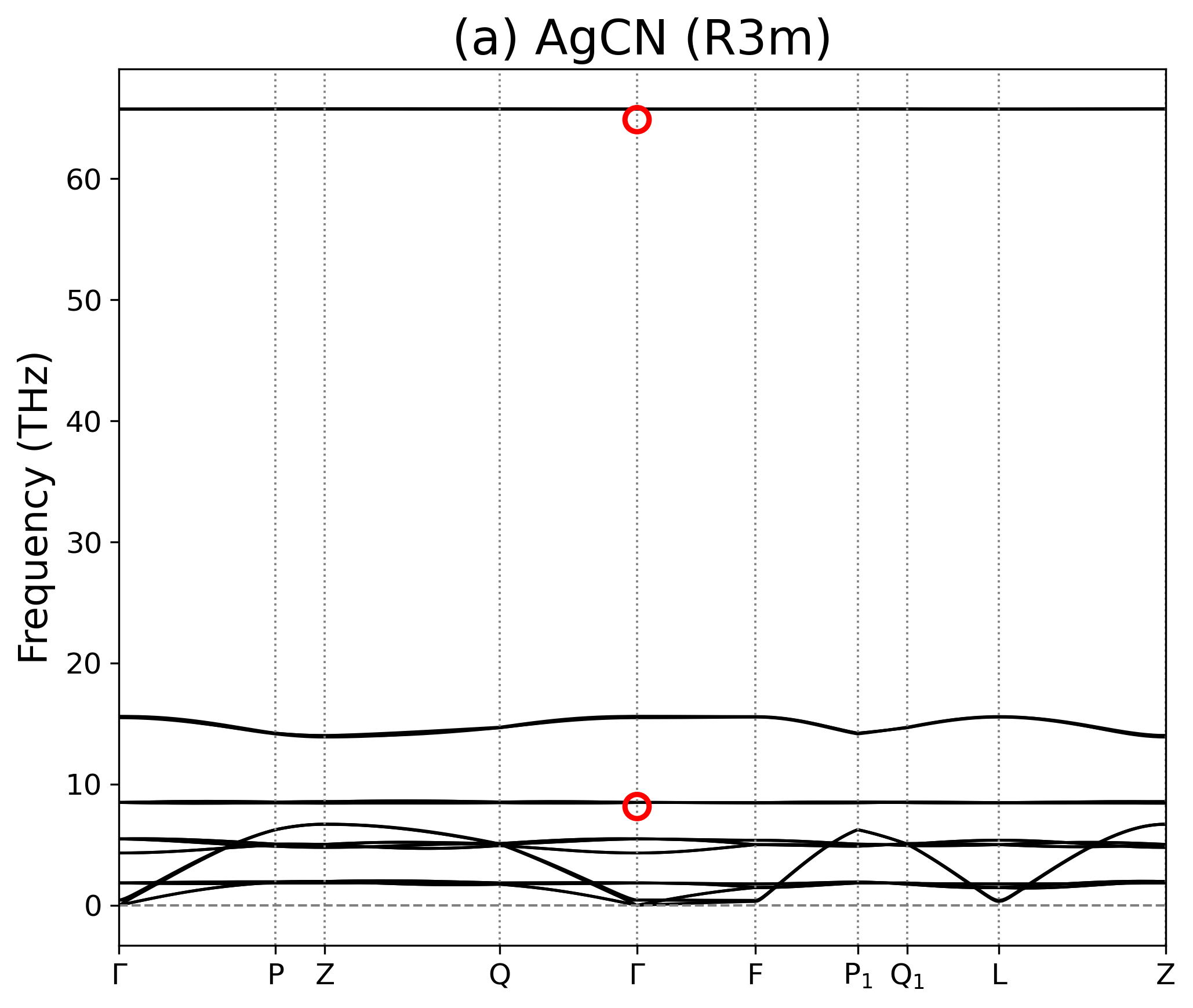}
            \put(58,36){%
                \includegraphics[width=0.3\textwidth]{R3m_phonon_path.png}
            }
            \put(-30,20){%
                \includegraphics[width=1.3\textwidth]{AgCN_phonon_displacements-removebg-preview.png}
            }
        \end{overpic}
        \label{fig:AgCN_phonons}
    \end{subfigure}
    \hfill
    \begin{subfigure}[b]{0.49\textwidth}
        \centering
        \begin{overpic}[width=\textwidth]{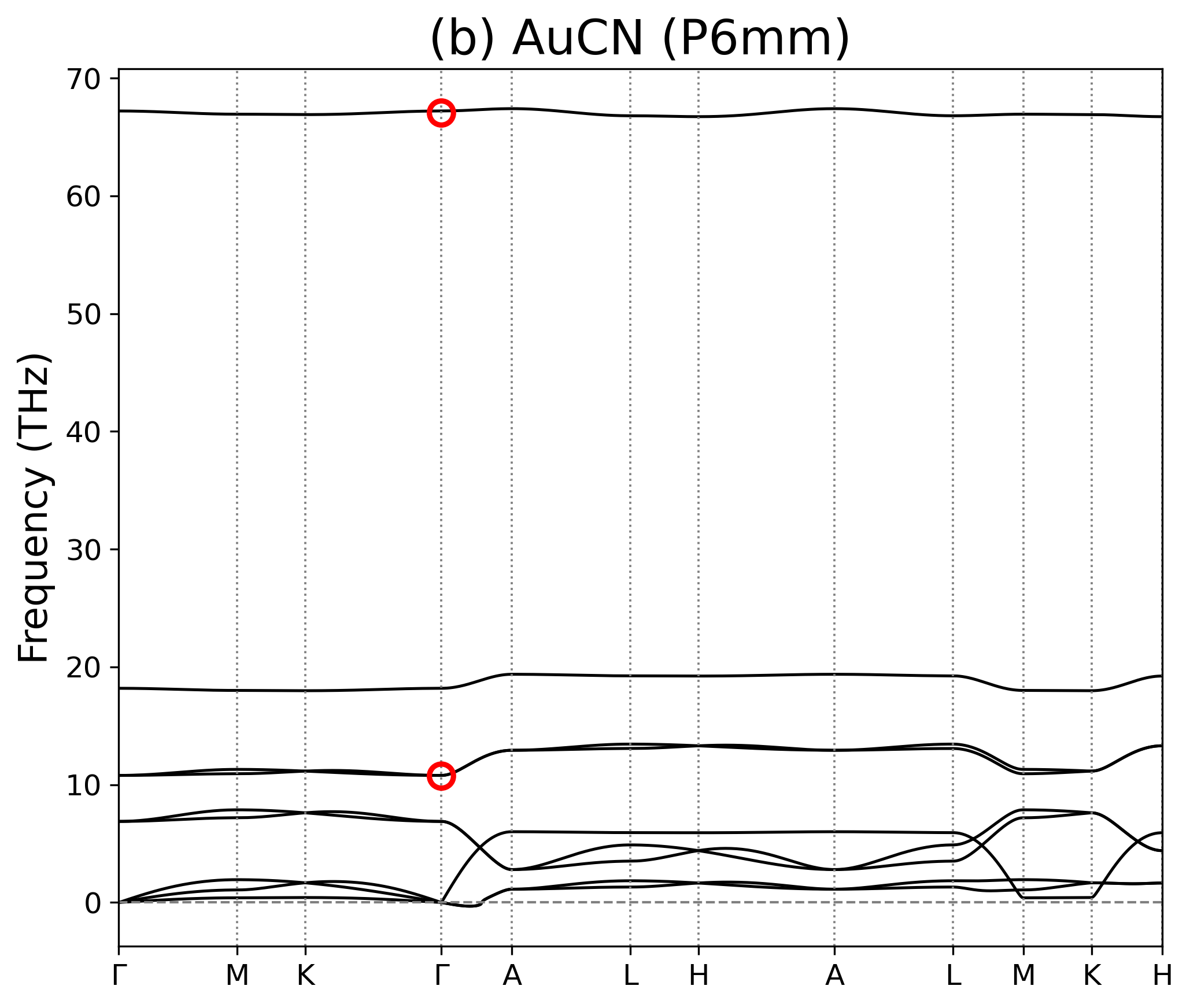}
            \put(57,45){%
                \includegraphics[width=0.4\textwidth]{P6mm_phonon_path.png}
            }
            \put(-35,26){%
                \includegraphics[width=1.3\textwidth]{AuCN_phonon_displacements-removebg-preview.png}
            }
        \end{overpic}
        \label{fig:AuCN_phonons_pbe}
    \end{subfigure}

    \caption{Phonon band structures of  the R3m phase of AgCN (a) and  the P6mm phase of AuCN (b) using PBE-D3. Insets on the right show the corresponding high-symmetry paths used in the phonon calculations (adapted from \cite{SETYAWAN2010299}). Insets on the left display the displacement vectors for the high-frequency modes. Red circles indicate the experimental high-frequency stretching and low-frequency bending modes from Ref.~\cite{Bowmaker1998}.
    }
    \label{fig:AgCN_AuCN_phonons}
\end{figure}

\begin{figure}[t]
\centering

\begin{subfigure}[b]{0.48\linewidth}
    \centering
    \includegraphics[width=\linewidth]{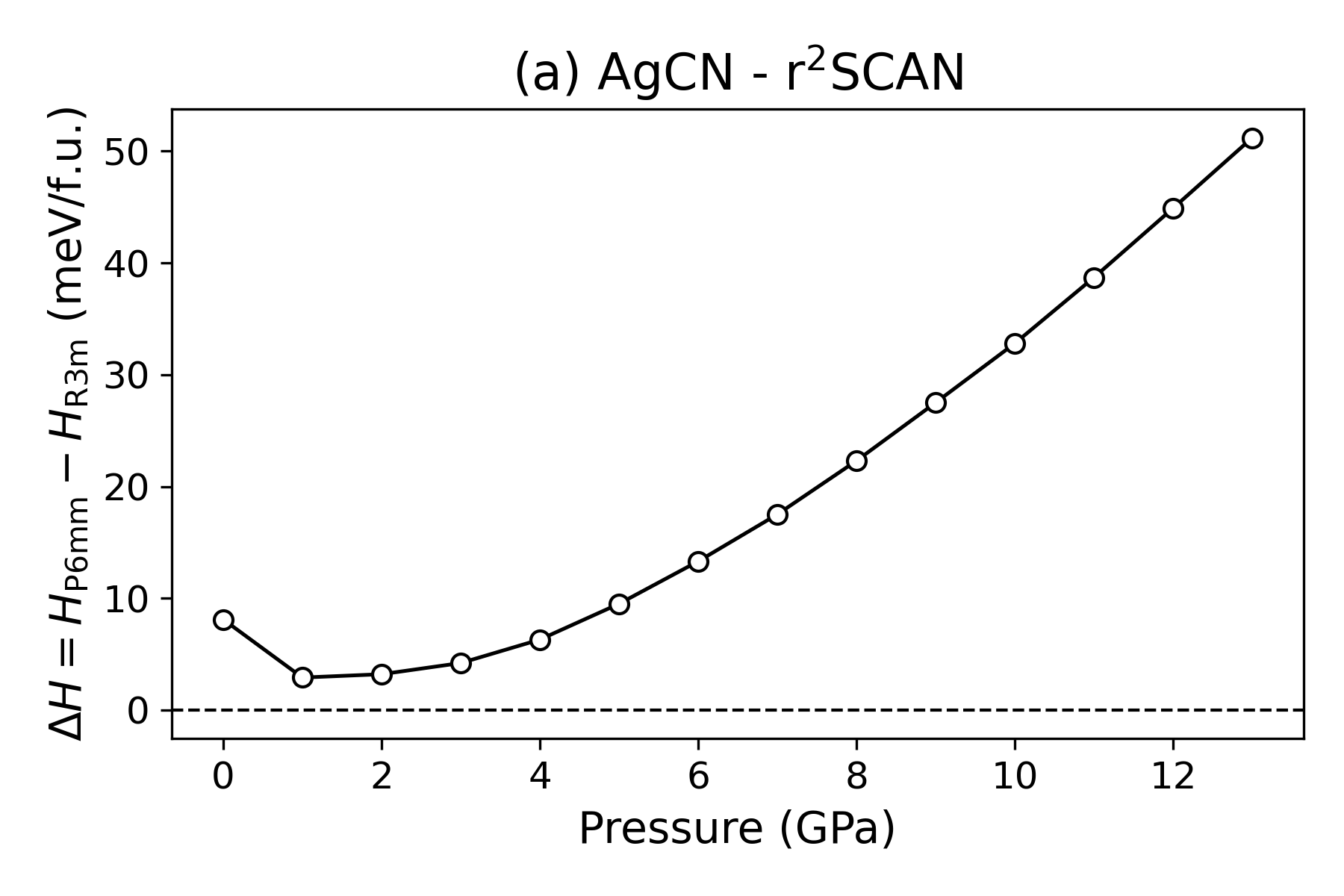}
\end{subfigure}
\hfill
\begin{subfigure}[b]{0.48\linewidth}
    \centering
    \includegraphics[width=\linewidth]{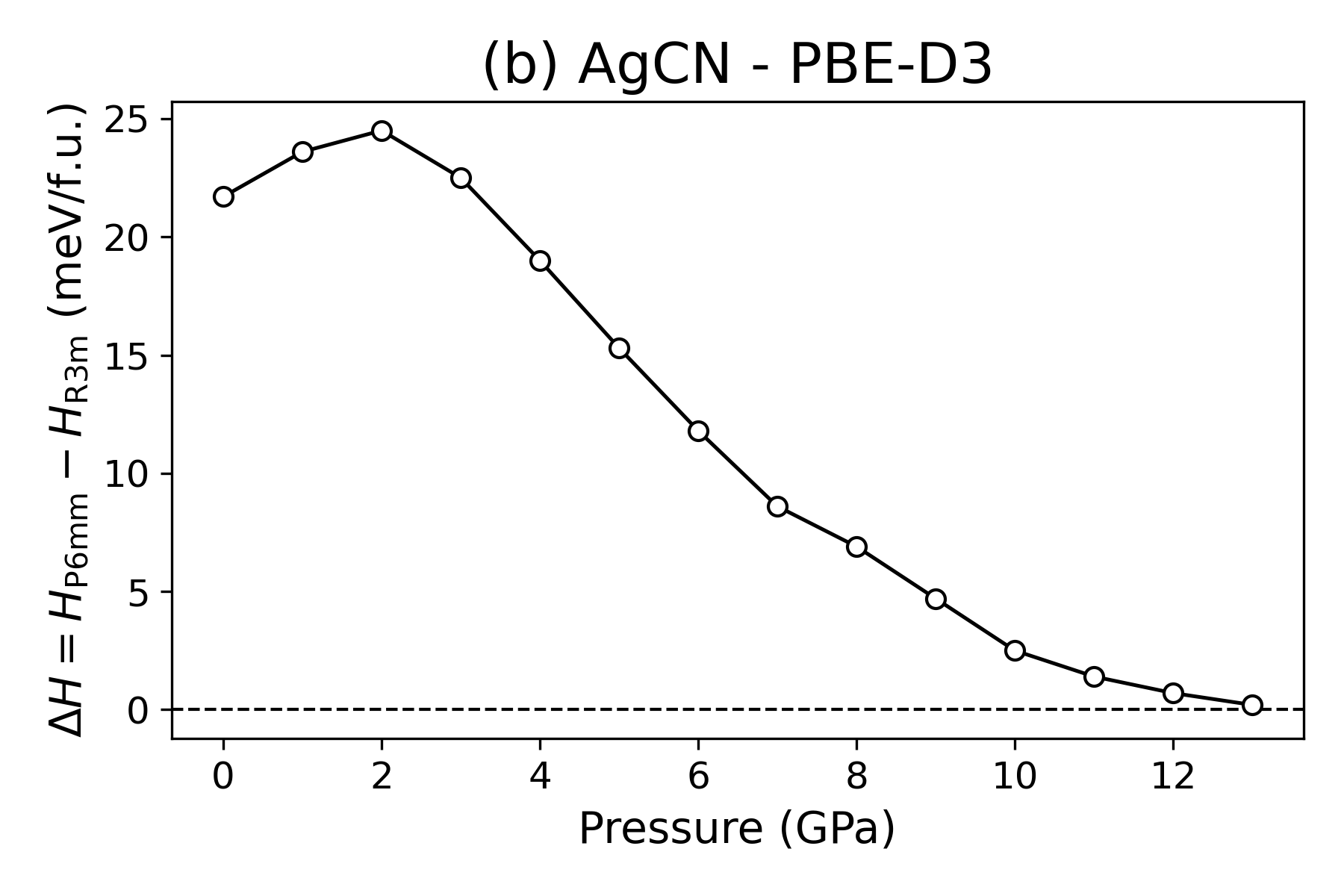}
\end{subfigure}

\vspace{0.3cm}

\begin{subfigure}[b]{0.48\linewidth}
    \centering
    \includegraphics[width=\linewidth]{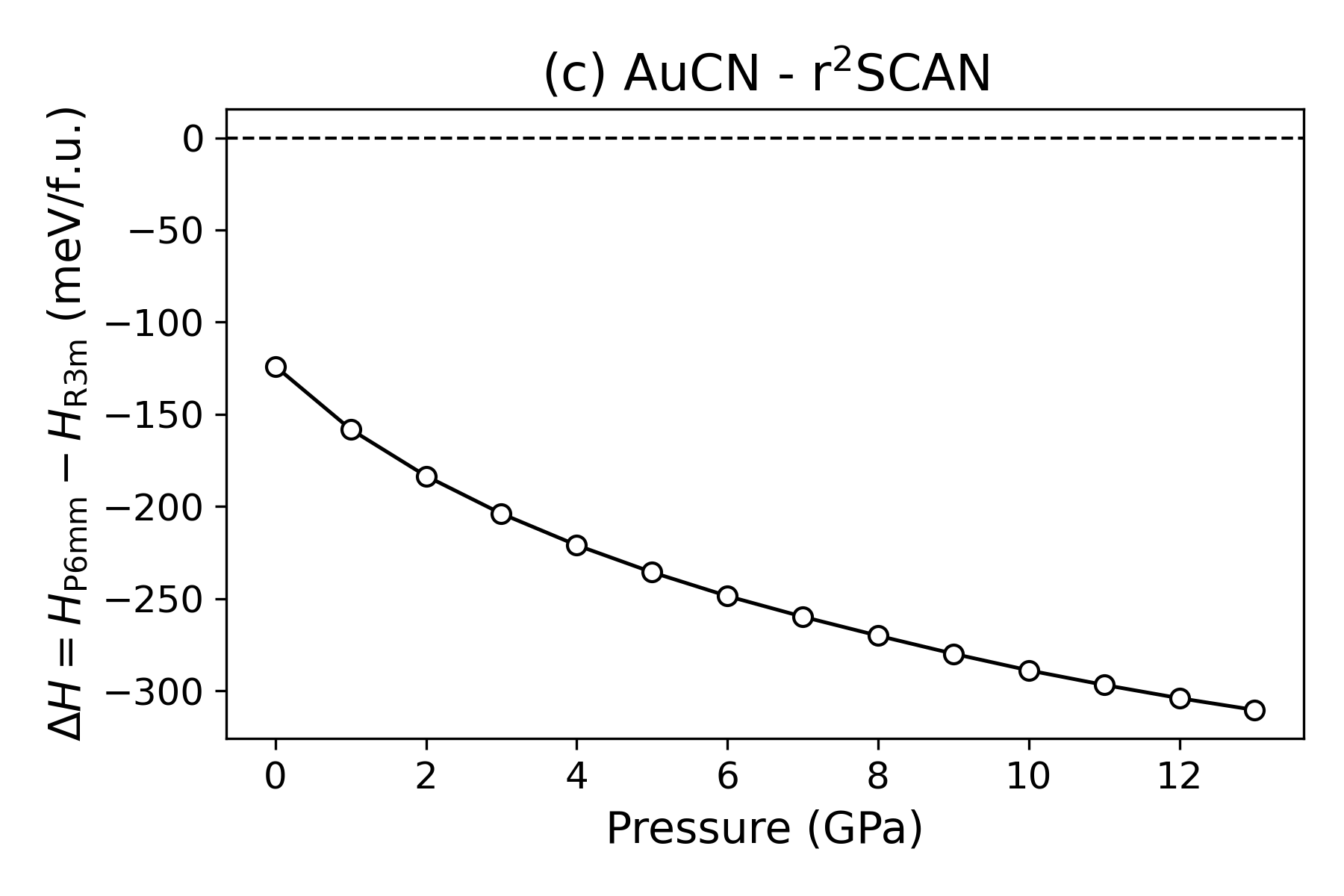}
\end{subfigure}
\hfill
\begin{subfigure}[b]{0.48\linewidth}
    \centering
    \includegraphics[width=\linewidth]{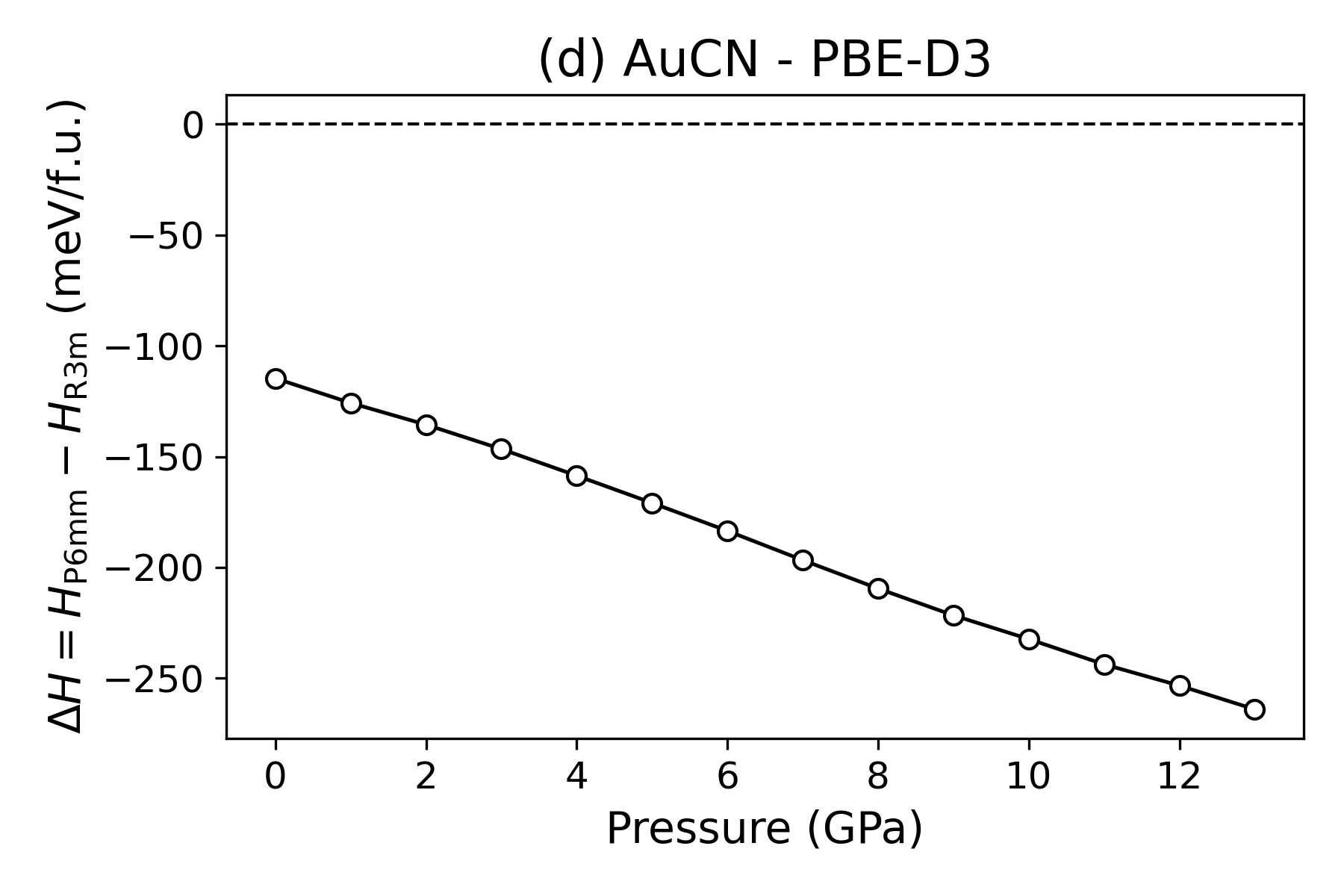}
\end{subfigure}

\vspace{0.3cm}

\begin{subfigure}[b]{0.48\linewidth}
    \centering
    \includegraphics[width=\linewidth]{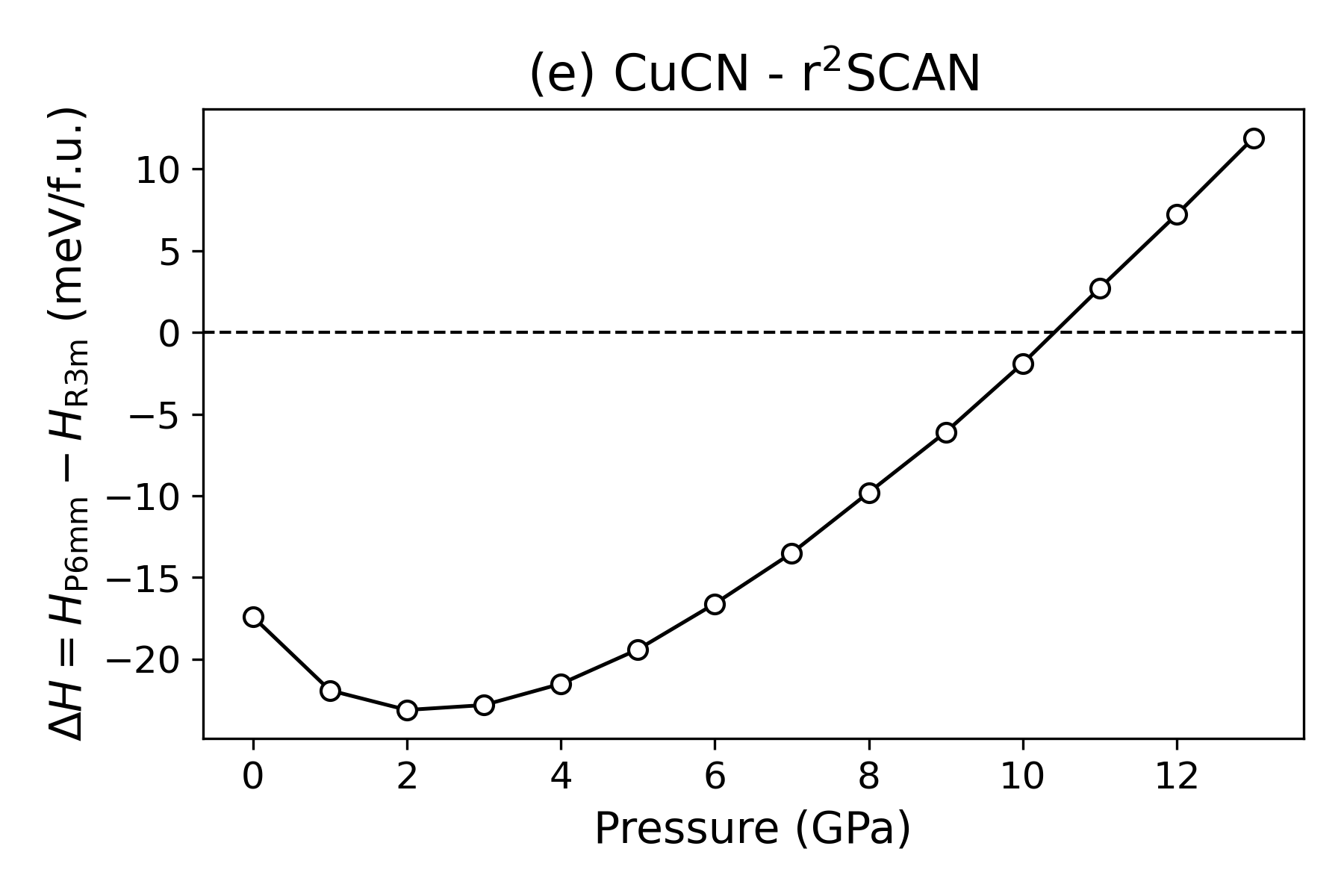}
\end{subfigure}
\hfill
\begin{subfigure}[b]{0.48\linewidth}
    \centering
    \includegraphics[width=\linewidth]{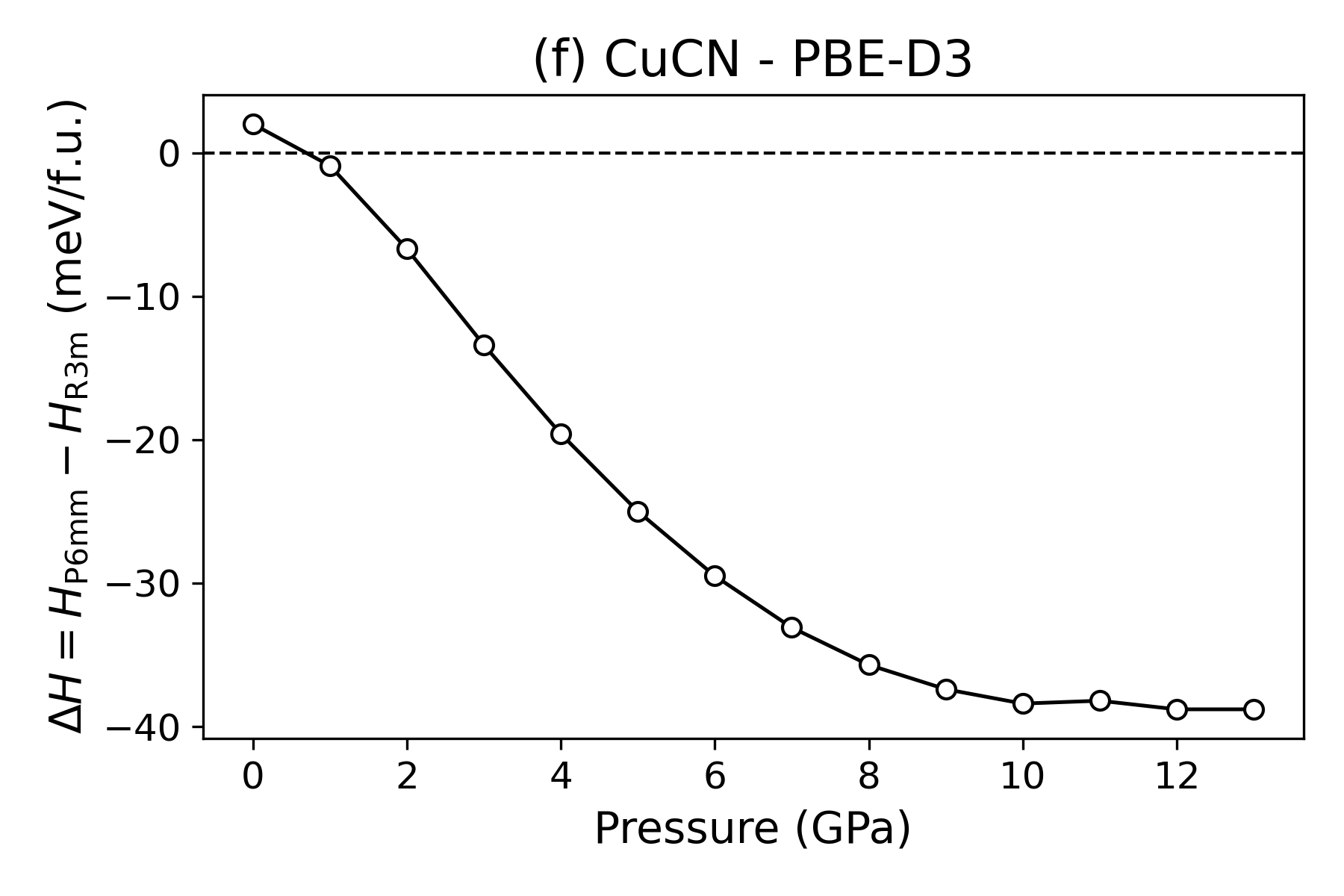}
\end{subfigure}

\caption{Difference in enthalpies between P6mm and R3m phases as a function of increasing pressure.}
\label{fig:enthalpy_diffs}

\end{figure}

\end{document}